%% file: main.tex
\documentclass[preprint,12pt]{elsarticle}
\usepackage[utf8]{inputenc}
\usepackage[T1]{fontenc}

\usepackage{amssymb}
\usepackage{amsmath}
\usepackage{miller}

\usepackage{lineno}

\usepackage{booktabs}
\usepackage{subcaption}

\biboptions{sort&compress}
\usepackage[colorlinks=true]{hyperref}

\journal{Computational Materials Science}

\begin{document}

\begin{frontmatter}


\title{A Self-Evolving Machine-Learning-Based Kinetic Monte Carlo Method for Modelling Thin-Film Growth} 

\author[label1,label2,label3]{Jyri Kimari}
\author[label2]{Flyura Djurabekova}
\author[label2,label1]{Kostas Sarakinos}
\affiliation[label1]{organization={Department of Physics, KTH Royal Institute of Technology},
            addressline={Roslagstullsbacken~21},
            postcode={114~21},
            city={Stockholm},
            country={Sweden}}

\affiliation[label2]{organization={Department of Physics and Helsinki Institute of Physics, University of Helsinki},
            addressline={P.O.~Box~43},
            postcode={FI-00014},
            city={Helsinki},
            country={Finland}}

\affiliation[label3]{organization={Institute of Technology, University of Tartu},
            addressline={Nooruse~1},
            postcode={50411},
            city={Tartu},
            country={Estonia}}

\begin{abstract}
  We present a kinetic Monte Carlo~(KMC) simulation framework parameterized by automatically sampling machine-learning~(ML) for modeling thin-film growth atom by atom. Given an interatomic potential energy function, the KMC algorithm builds an ML-based regression model for rate parameters on runtime, being trained on the local atomic environments encountered during the system evolution. New environments are continuously added to the training set in a self-evolving manner at points where the ML model estimates high uncertainty. As the simulation progresses, the ML model gains confidence, and the quick estimation of rates increasingly overtakes the relatively-expensive nudged elastic band calculations, promoting computational efficiency while retaining high fidelity description of the atomic diffusion kinetics. As a test case, we simulate the sub-monolayer growth of Ag on Ag~\hkl{111}, where we demonstrate adatom islands forming in shapes and densities in accordance with the underlying atomistic interaction model, the theoretical framework, and available experimental results related to thin-film nucleation and growth.
\end{abstract}



\end{frontmatter}



\section{Introduction}
\label{sec:introduction}

  Theory-informed design and synthesis of thin films and nanoscale materials requires computational modeling of surface and bulk diffusion processes with atomic-scale accuracy at  experimentally-relevant timescales. State-of-the-art methods cannot meet those two requirements simultaneously. Molecular Dynamics~(MD) provides access to a detailed deterministic description of atomic-scale processes, but at timescales that do not exceed a few tens of nanoseconds~\cite{lesar2013introduction}. This is because atomic diffusion is a rare event that occurs at frequencies many orders of magnitude smaller than that of thermal vibration of atoms, thereby limiting the length of the timestep required for accurate integration of the equations of motion. Kinetic Monte Carlo~(KMC) models forego the description of atomic vibrations allowing for the simulation timescale to be extended to a few seconds and beyond~\cite{voter2007introduction}. However, this comes at the expense of atomistic accuracy and completeness as KMC models are parameterized directly in terms of the activation barriers, which have a complex dependence on the local atomic environment~(LAE) of the diffusion events. Due to this complexity, full tabulation of every barrier is practically impossible. One common approximation to overcome this inherent weakness of KMC is the use of bond-counting schemes~(BCS), where the description of the LAE is simplified down to the number of first- and second-nearest neighbour bonds that the diffusing atom breaks and forms along its migration path~\cite{michely2004islands}. The activation energy is thus a function of the bonds, and the coefficients of this function are fitted to a sample of representative barriers obtained, e.g., by molecular statics calculations.

  BCS parameterizations have been used to simulate deposition and diffusion on solid surfaces already before Bortz, Kalos and Lebowitz introduced the nowadays standard $n$-fold KMC algorithm in~1975~\cite{bortz1975new}. Abraham and White modeled adsorption, evaporation and migration on cubic close-packed surface in~1970, describing all event rates in terms of single bond energies~\cite{abraham1970computer}. Gilmer and Bennema carried out a similar study on crystal growth from solution in~1972~\cite{gilmer1972simulation}. The first efforts in developing a BCS approach for \emph{specific} materials, instead of generic crystals, were taken by Voter in~1986 for the case of Rh clusters diffusing on the Rh~\hkl{100} surface~\cite{voter1986classically}. Since then, the development of BCS-reliant KMC and lattice gas~(LG) models has branched both in the simple, generic-system~\cite{clarke1987origin,clarke1988growth,schroeder1997bond,combe2000changing,vladimirova2000new,kotrla2000submonolayer,kotrla2000submonolayera,kotrla2001effects,yang2018three} and material-specific~\cite{clarke1991theory,ratsch1994scaling,jacobsen1995island,jacobsen1996island,breeman1996computer,merikoski1997effect,smilauer1998activated,biham1998models,mehl1999models,fichthorn2000island,kohler2000investigation,bhuiyan2011kinetic,jansson2016long,jansson2020tungsten,garcia2024dft} directions. The BCS approach was also taken in Refs.~\cite{lu2018formation,gervilla2019dynamics,gervilla2020coalescence} to simulate Ag deposition and diffusion on Ag~\hkl{111} and generic weakly-interacting substrates with the purpose of studying the dynamics of three-dimensional morphological evolution of thin films.

  Even after almost~40 years, Voter's vision of using simple, or at least tractable, models to capture ``classically accurate'' dynamics at the fraction of the cost of MD simulations, has not been fully realized. Instead, it was understood rather early on~\cite{schroeder1997bond} that the number of necessary parameters to accurately represent the microscopic evolution of real-material surfaces in KMC is effectively infinite. This realization, along with the increasing computational power at researchers' disposal, has spawned the class of on-the-fly KMC models that are launched without an \textit{a priori} list of events and associated barriers. Instead barriers are calculated on runtime only for the environments they encounter, using, e.g., saddle point search for identifying diffusion events~\cite{barkema1996event,henkelman2003multiple}. For computational efficiency, any found events and barriers may be saved in a database for later reuse~\cite{trushin2005self,rahman2004cluster,el2008kinetic,konwar2011off,beland2011kinetic,latz2012three}. Concurrently, the so-called ``hybrid'' KMC methods, where the system is treated partially by the stochastic event-based algorithm and partially by deterministic MD, have shown some popularity~\cite{zoontjens2007hybrid,zhu2017hybrid,olowookere2025integrated}.

  The vast majority of thin-film applications concerns the deposition of multi-component layers on chemically heterogeneous substrates. The implementation of BCS-based models in such growth scenarios is challenging as film-forming species are expected to encounter numerous and \textit{a priori} unknown LAEs, each of which may require a separate BCS function for barrier calculation. These limitations have been circumvented chiefly by either limiting the growth to two-dimensional, submonolayer regime~\cite{smilauer1998activated,kotrla2000submonolayer,kotrla2000submonolayera,kotrla2001effects,cougnon2018impurity}, or (in the case of three-dimensional growth) by coarse-graining sites to represent multiple atoms~\cite{bouaouina2018nanocolumnar} and by imposing a rigid NaCl~\cite{antoshchenkova2012kinetic,nita2016three,bouaouina2018nanocolumnar,mareus2020study,mareus2021effect,maleki2025modeling} or wurtzite~\cite{kaufmann2016critical,chugh2017lattice} lattice structure with predetermined sites for different atomic species. Additionally, the diffusion of impurity species~\cite{cougnon2018impurity}, or even all deposited atoms~\cite{karewar2024minimum} may also be restricted. In some cases, hybrid KMC-MD approaches have been applied to molecular deposition of thin films~\cite{abbott2022kinetically,ntioudis2023hybrid}.

  To our knowledge, there is no model to simulate with full atomistic accuracy growth of multi-elemental films in the multilayer growth regime where the LAEs cannot be predicted \textit{a priori}. This includes the growth of metals and alloys on weakly-interacting substrates which exhibit a pronounced~3D morphological evolution. These systems are relevant for applications such as photovoltaics~\cite{wang2014transparent}, energy-saving windows~\cite{dalapati2018transparent}, and biosensors~\cite{zhao2018sensitivity}. Specific challenges also include facet formation and large anisotropy in the diffusivity of the film-forming species. The growth of new facets with off-lattice sites was captured by L\"u et al. using a BCS model~\cite{lu2018formation}, but attempts to extend this approach to the multi-elemental case have been unsuccessful.

  Breakthroughs in machine-learning~(ML) methods in the last two decades have paved the way for development of new, efficient and accurate atom-level simulation techniques. Bart\'ok et al. pioneered the Gaussian approximation potential~(GAP) approach~\cite{bartok2010gaussian} and the smooth overlapping atomic positions~(SOAP) atomic environment descriptors~\cite{bartok2013representing}. ML interatomic potentials have since been well-established in MD simulations~\cite{deringer2019machine,mortazavi2023atomistic}. Djurabekova, Castin and others have applied artificial neural networks to KMC simulations of bulk diffusion in Fe-based alloys in a rigid lattice~\cite{djurabekova2007artificial,castin2008use,castin2009prediction,castin2009modelling,castin2010calculation,soisson2010atomistic,castin2011atomistic,castin2011modeling,pascuet2011stability,castin2012mobility,messina2017introducing}, correctly capturing the migration and aggregation of vacancies and impurities. Taking inspiration from Bart\'ok's work, Castin et al. applied this methodology also to the lattice-free case, simulating vacancy migration near grain boundaries~\cite{castin2014predicting}. In another work, they trained a neural network potential on \textit{ab initio} data and used it to parameterize binary lattice-free Fe-Cu and Fe-Cr KMC models~\cite{castin2017improved}. ML parameterizations for surface diffusion KMC face unique challenges pertaining to sampling training data from the highly heterogeneous space of LAEs with large variance in migration barriers, the presence of unstable atomic environments (overhang positions), and difficult-to-interpret migration events with intermediate potential energy local minima~\cite{baibuz2018migration,kimari2020application}. Limited (two-dimensional fcc~\hkl{100}) cases have been studied by Sastry et al.~\cite{sastry2005genetic} by genetic programming and Verma et al.~\cite{verma2013cluster} by a cluster expansion approach.

  In the present work, we develop a novel, ML augmented kinetic Monte Carlo~(ML-KMC) tool for modeling deposition, diffusion, and growth of metal films on fcc surfaces. The model overcomes the challenges pertaining to intermediate energy minima along the migration paths and overhang positions by allowing hops to multiple distances, i.e., landing in those intermediate minima or diffusion over step-edges. Lattice sites for new facets at arbitrary angles and hexagonal close-packed~(hcp) stacking faults are freely generated as necessary. The atomic diffusion rates are learned on-the-fly, yielding an efficient and accurate description of kinetics without requiring any \textit{a priori} knowledge of the likely diffusion events or any particular sampling scheme of the training data. We demonstrate the feasibility of our approach by simulating Ag homoepitaxy on the~\hkl{111} surface and find that the model correctly captures island nucleation density and shape evolution at different temperatures. Concurrently, the model can be readily extended to multiple chemical elements and hence can be viewed as the first crucial step towards Voter's vision of fast and accurate modeling of surface dynamics of multicomponent thin films on chemically heterogeneous substrates. 

  The remainder of the article is organized as follows. In section~\ref{sec:methods} we describe the ML-KMC algorithm and the physics it captures, and in section~\ref{sec:results_and_discussion} we present the results and discussion for the simulations of Ag homoepitaxial growth by our model. Conclusions are drawn in section~\ref{sec:conclusions}.

\section{Methods}
\label{sec:methods}

  The computational tool described herein is a self-parameterizing, ML augmented KMC model for metal vapor deposition and diffusion on a metal substrate and subsequent film growth evolution. The currently implemented substrate type is the fcc~\hkl{111} surface, but extension to other crystal structures and orientations can be done without major modifications to the crucial parts of the program. The model is tested for homoepitaxial Ag deposition, but the code can be readily run using any element or alloy for which an interatomic potential is available in the Large-scale Atomic/Molecular Massively Parallel Simulator~(LAMMPS) that is used for calculation of migration barriers.~\cite{plimpton1995fast}

  The main simulation flow follows the traditional $n$-fold KMC algorithm~\cite{bortz1975new}. Each step, an event is chosen at random from among all jump and deposition events available at that step. The probability of an event to be chosen is proportional to its rate. Deposition takes place at a constant, user-given rate, while jump rates $\Gamma$ are calculated via the Arrhenius equation
  \begin{equation}
    \label{eq:arrhenius}
    \Gamma = \nu \exp\left(\frac{-E_\mathrm{m}}{k_\mathrm{B}T}\right),
  \end{equation}
  where $\nu$ is the pre-exponential attempt frequency factor, $E_\mathrm{m}$ is the migration energy barrier of the jump, $k_\mathrm{B}$ is the Boltzmann constant, and $T$ is the temperature of the system. After the event, the simulation time is incremented according to the residence-time algorithm
  \begin{equation}
    t' = t + \frac{-\log u}{\sum_i \Gamma_i},
  \end{equation}
  where $u$ is a uniform random number, and $\Gamma_i$ is the rate of the $i$th event. The sum includes all available events in the system.

\begin{figure}
    \centering
    \includegraphics[width=0.9\linewidth]{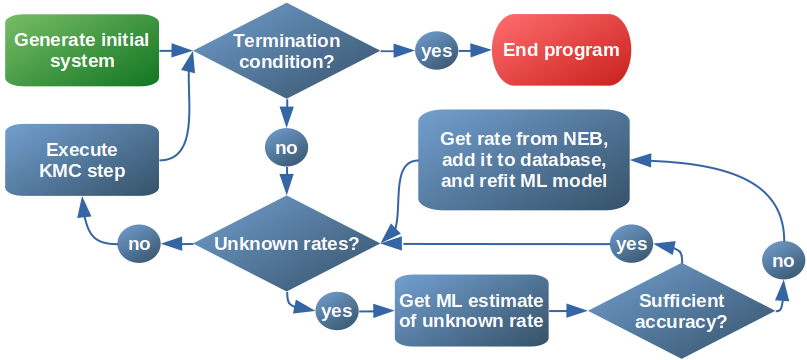}
    \caption{Simplified flowchart of the ML-KMC program. The termination condition is either a maximum number of KMC steps, a maximum simulated time $t$, or a maximum elapsed wall time. Execution of a KMC step encompasses choosing and executing a deposition or a diffusion event, and incrementing of $t$. Acceptance of the machine-learning~(ML) rate (barrier) estimate is decided based on the reported uncertainty and a user-defined uncertainty threshold. The nudged elastic band~(NEB) calculation includes first a check in the ML database for an exact match of the given LAE and then a conjugate-gradient minimization of the final configuration of the jump.}
    \label{fig:flowchart}
  \end{figure}

  The migration barriers are, at first, calculated on-the-fly via the nudged elastic band~(NEB) method~\cite{mills1994quantum,mills1995reversible} implemented in the LAMMPS software. Concurrently, a Gaussian process regression~(GPR)~\cite{rasmussen2005gaussian} ML model is trained with the migration barrier data. The GPR implementation is taken from the GAP library~\cite{bartok2010gaussian}, distributed as part of the Quantum mechanics and interatomic potentials~(QUIP) package~\cite{csanyi2007expressive}. As input for GPR, the smooth overlap of atomic positions~(SOAP) descriptor~\cite{bartok2013representing} is used for LAE of a jump event up to the fourth-nearest neighbour~(4nn) distance. The SOAP and GPR parameters are given in Tab.~\ref{tab:soap_gpr}. The GPR outputs an estimate of the barrier corresponding to a given LAE, as well as an estimate of the associated uncertainty of the barrier. Whenever the uncertainty exceeds a user-defined threshold, a complete NEB calculation is launched to add a new data point to the training set. As the GPR model grows more confident, the frequency of NEB calculations decreases, increasing the computational efficiency. The workflow of the algorithm is illustrated in Fig.~\ref{fig:flowchart}.
  \begin{table}[ht]
    \centering
    \caption{Parameters of the SOAP descriptor and the GPR fitting and calling. We refer the reader to the GAP and SOAP documentation for the description of the parameters. The uncertainty tolerance is the threshold for accepting or rejecting the barriers returned by the GPR on runtime.}
    \label{tab:soap_gpr}
    \begin{tabular}{lll}
    \toprule
        Parameter                  & Value         \\
    \midrule
        SOAP cutoff                & 11.5109\,\AA \\
        $l_\mathrm{max}$           & 3            \\
        $n_\mathrm{max}$           & 4            \\
        Cutoff transition width    & 0.5\,\AA     \\
        Atomic Gaussian width      & 1.0          \\
        Kernel $\zeta$ coefficient & 20           \\
        Signal variance $\delta$   & 20.0\,eV     \\
        $\sigma_y$                 & 0.1\,eV      \\
        Covariance type            & dot product  \\
        Regularisation             & 0.01 eV      \\
        Uncertainty tolerance      & 0.05 eV      \\
    \bottomrule
    \end{tabular}
  \end{table}

  We maximize the consistency of the ML model training set to increase the robustness of learning and pattern recognition by only including barriers that correspond to jumps between two \emph{true potential energy minima, separated by a single barrier}, that are furthermore calculated purely by NEB using the underlying potential energy function \emph{without any adjustments}. In other words, we intentionally \emph{avoid} the following three cases:
  \begin{enumerate}
      \item Events with artificially stabilised initial or final configurations, such as jumps to overhang positions on island edges. Resolving barriers for such events would require additional minimization constraints known as the \emph{tethering} technique~\cite{jansson2016long,baibuz2018migration}.
      \item Events with intermediate local minima of the potential energy. These include the simple adatom hop between two fcc sites on the~\hkl{111} surface, which takes place through a metastable hcp site.
      \item Barriers calculated or adjusted by an analytic formula. Such formulae are another approach that can be used to assign barriers for low-coordinated configurations~\cite{jansson2016long}, or to destabilise off-lattice positions, as was done in Ref.~\cite{lu2018formation} for hcp sites on the~\hkl{111} surface.
  \end{enumerate}
  To satisfy these conditions, we permit atoms to jump to both shorter than the fcc~1nn distance, to capture fcc-hcp jumps, and to longer than the~1nn distance, to allow island edge crossing without artificial barriers to overhang positions. This approach necessitates a \emph{refined lattice} system with fcc and hcp sites being dynamically generated on all~\hkl{111} surfaces and facets. We briefly describe our site-generation algorithm in the following.
  
\begin{figure}[ht]
    \centering
    \includegraphics[width=0.6\linewidth]{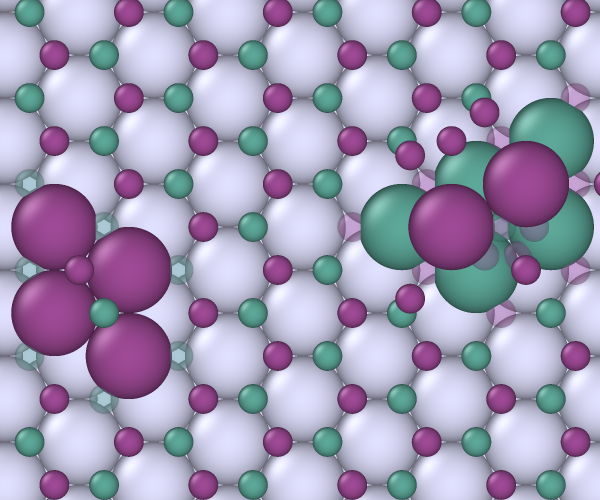}
    \caption{Top view of the Ag~\hkl{111} surface slab with eleven adatoms deposited on top. Substrate atoms are silver colour, adsorption sites are small spheres. Deposited atoms and the adsorption sites are coloured green if they conform to the underlying fcc structure, and purple if they are at off-lattice hcp stacking fault locations. New adsorption sites have been generated to three- and four-coordinated sites above the deposited atoms. Adsorption sites that are blocked by nearby atoms are marked as transparent.}
    \label{fig:site_illustration}
  \end{figure}

  In the beginning of the simulation, the substrate slab (periodic in the $x$ and $y$ directions and an open surface in the $z$ direction) of user-defined dimensions is generated. Next, a layer of initial adsorption sites is added on top of the substrate. This layer includes all available fcc and hcp sites on the substrate surface. Whenever a new atom jumps to or is deposited at a site adjacent to at least two other sites occupied by non-substrate atoms (i.e., a triplet forms after the event), a check is made for generation of new adsorption sites. New sites are generated so that they are at~1nn distance from all three atoms in the triplet. Utilizing the neighbour-lists and the fixed interlayer distances in the fcc lattice, the step of generating a new site is very efficient and does not require a square root function at any point. Growing the lattice this way allows the organic formation of new facets oriented at arbitrary angles to the original substrate. An illustration of newly-generated sites on top of deposited atoms is shown in Fig.~\ref{fig:site_illustration}.

\begin{figure}[ht]
    \centering
    \includegraphics[width=0.6\linewidth]{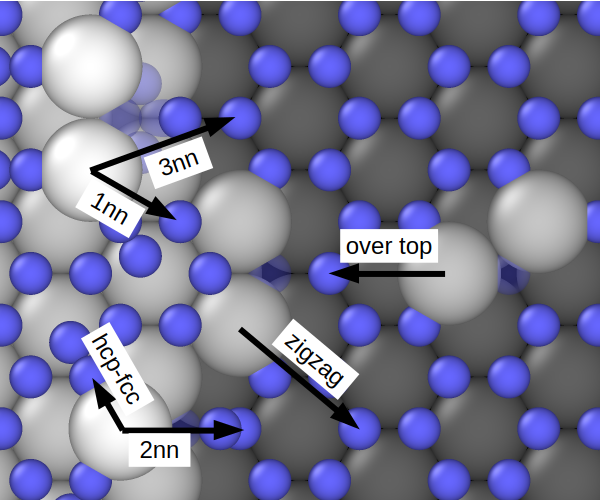}      
    \caption{The different jump lengths allowed in the system.}
    \label{fig:allowed_jumps}
  \end{figure}

  We allow six different hop distances to up to~3nn adsorption sites from the jumping atom (see Tab.~\ref{tab:jump_shells} and Fig.~\ref{fig:allowed_jumps}). In addition to hops, exchange-displacement events involving two~1nn atoms and one adsorption site at the~1nn distance to either of them are permitted.
  
\begin{figure}
      \centering
      \begin{subfigure}{0.49\linewidth}
        \includegraphics[width=\linewidth]{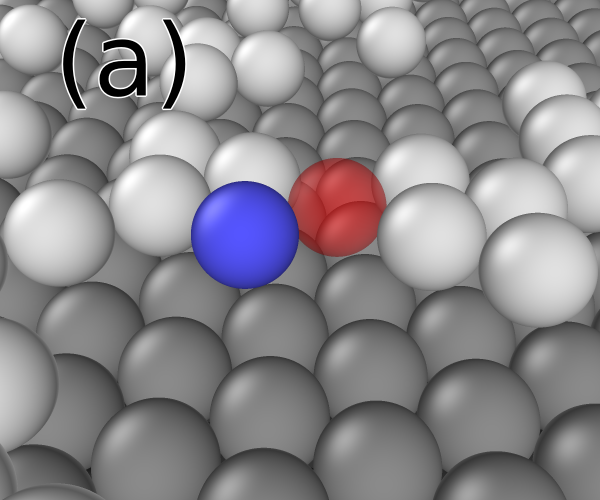}          
      \end{subfigure}
      \begin{subfigure}{0.49\linewidth}
        \includegraphics[width=\linewidth]{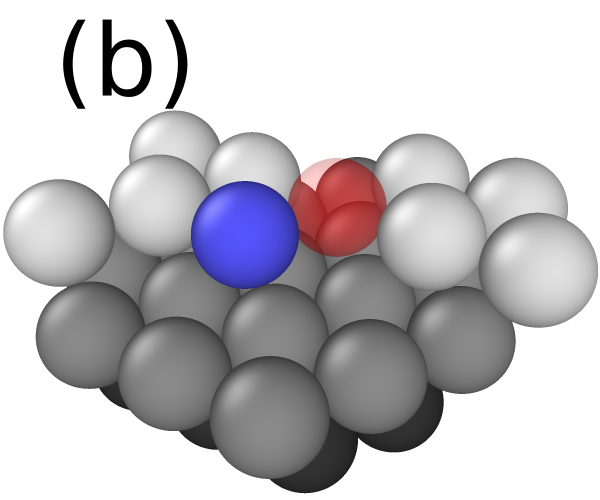}          
      \end{subfigure}
      \caption{Local atomic environment~(LAE) of the jump of the blue-coloured atom to the adsorption site marked by a transparent red sphere. Surrounding atoms are coloured in grayscale according to their $z$ coordinate. \emph{(a)} The full environment of the jump. \emph{(b)} The extracted LAE up to the~4nn distance of the initial and the final position, to be input to the SOAP descriptor forming function.}
      \label{fig:LAE}
  \end{figure}

  \begin{table}[ht]
    \centering
    \caption{The six jumping distances allowed in the simulations. See Fig.~\ref{fig:allowed_jumps}a for visual descriptions of the jumps.}
    \label{tab:jump_shells}
    \begin{tabular}{rl}
      \toprule
      Distance & Description \\
      \midrule
      $\sqrt{1/6}\,a_0$ & \hkl{111} fcc-hcp jump\\
      $\sqrt{1/2}\,a_0$ & 1nn jump \\
      $\sqrt{2/3}\,a_0$ & Jump over top site \\
      $a_0$ & 2nn jump \\
      $\sqrt{7/6}\,a_0$ & \hkl{111} ``zigzag'' jump \\
      $\sqrt{3/2}\,a_0$ & 3nn jump \\
      \bottomrule
    \end{tabular}
  \end{table}
  The dependence on the direction of the jump is introduced in the ML model by including the atoms around the final position of the jump in the SOAP descriptor---the descriptor that is developed about the jumping atom ``sees'' the atomic density further in the direction of the jump. Different ML models are fitted for each of the possible jump lengths in the system. This approach reduces the heterogeneity of the training data and more rigorously enforces the dependence on the jump length. See Fig.~\ref{fig:LAE} for an illustration of the LAE.

  To further improve the ML performance by reducing the heterogeneity of the training set, the data is also split according to the coordination of the jumping atom at the initial and the final positions (and, for exchange processes, the coordination of the middle atom as well). Separate ML models are trained for each combination of coordination numbers.

  Since NEB calculations are the (initial) computational bottleneck in the simulations, they are avoided when possible. When the result of an ML call indicates that a NEB barrier is required for a given LAE, the ML training database is first searched for an exact match, so no calculation is repeated. In the case of no match, a conjugate-gradient minimization is first performed to the final state of the process---a single minimization is faster than full NEB. If the final state relaxes improperly (either back to the initial state or to a third, unrelated state), no NEB is carried out for this jump and the process is marked forbidden. Only if minimization is successful, a NEB calculation is launched, the returned barrier used in the KMC rate calculation, and the backward and forward barriers added to the ML training data.

  The code also implements a rudimentary pattern recognition scheme to avoid unnecessary minimizations. Whenever a minimization fails, the associated process is marked as forbidden, and no barrier is calculated or saved in the ML training data. Hence, the next time this process rate is requested, a new minimization would be launched. To avoid this, the minimization results (success/fail) are saved in another database, indexed by ``fingerprints'' that are generated about the position of the jumping atom in the final state. The fingerprint is an array of integers that count the number of neighbours in each shell around the jumping atom.

  Because of the relatively high mobility of terrace adatoms on the fcc~\hkl{111} surface (activation barrier in Ag being around~100\,meV), deposition at a low rate (<100 monolayers/s) at room temperature or higher is likely to be computationally inefficient: e.g., depositing~10 monolayers per second at~300\,K on a substrate of $n_\mathrm{monolayer}=10\,000$ atoms per monolayer gives approximately one deposition event per~6~million KMC diffusion jumps of an adatom. As more atoms are deposited and form islands, the problem is alleviated because the time an adatom spends on the surface is limited by the rate of encountering an island edge and becoming captured. To speed up the initial stages of the simulation, the jump rates are slowly lowered while no deposition events take place. Specifically, if the system is trapped in a state with no newly deposited atoms for $10\times n_\mathrm{monolayer}$ KMC steps, all jump rates are divided by~2. After another $10\times n_\mathrm{monolayer}$ steps without new depositions, the rates are again halved, and so on. Whenever a new atom is deposited, all jump rates are restored to their original values. Scaling with $n_\mathrm{monolayer}$ is used to account for the fact that the lifetime of the monomer increases with substrate size and thereby to avoid improper (either too fast or too slow) rate modification. The rate modification scheme doubles as a solution to the so-called ``low-barrier'' or ``flickering'' problem, where system evolution is stalled by back-and-forth repeated processes that are dominantly selected in the $n$-fold algorithm: when flickering occurs, each newly-deposited adatom that attaches near the environment of the flickering process gives an opportunity for the situation to resolve by the system finding a more stable configuration. This approach is conceptually similar to Chatterjee and Voter's accelerated superbasin KMC~(AS-KMC)~\cite{chatterjee2010accurate}, except that the ``basins'' are considered to be separated only by new atom depositions, rather than higher-barrier jumps. A rigorous implementation of AS-KMC would be memory-wise very expensive due to the bookkeeping of visited states in the continuously evolving dynamic lattice system.

  The underlying potential energy function used in the NEB barrier calculations is a Finnis-Sinclair type ternary Cu-Ag-Au potential by Ackland and Vitek~\cite{ackland1990manybody}. Eleven replicas are used in the NEB calculations, with cross-replica force constant equal to~1.0\,eV/\AA$^2$.

  The simulations in the present article are performed in a simulation cell of~10\,044 fcc sites (270.23\,\AA $\times$ 268.70\,\AA) with periodic boundary conditions in the lateral dimensions. Atoms are deposited on an initial substrate of~6 monolayers of fcc Ag crystal. The substrate atoms are fixed in the KMC events (they cannot directly participate in jumps), but they are allowed to freely relax during NEB barrier calculations, apart from the very bottom layer, which represents a semi-infinite piece of bulk Ag.

  The source code for the ML-KMC program is available at GitHub~\cite{kimari2026mlkmcrepo}, licensed under GNU General Public License version~3.

\section{Results and discussion}
\label{sec:results_and_discussion}

\subsection{Initial training and ML model evaluation}

  For the initial training of the barrier prediction model, a sequence of deposition simulations was run, starting at temperature of~500\,K for~2 days of wall time, and subsequently at~400\,K to~100\,K at increments of~100\,K for~3 days of wall time at each temperature. The deposition rate was~10 monolayers/s. By the end of this training pipeline, a total of~10\,291 NEB calculations had been carried out.

\begin{figure}[ht]
      \centering
      \includegraphics[width=0.7\linewidth]{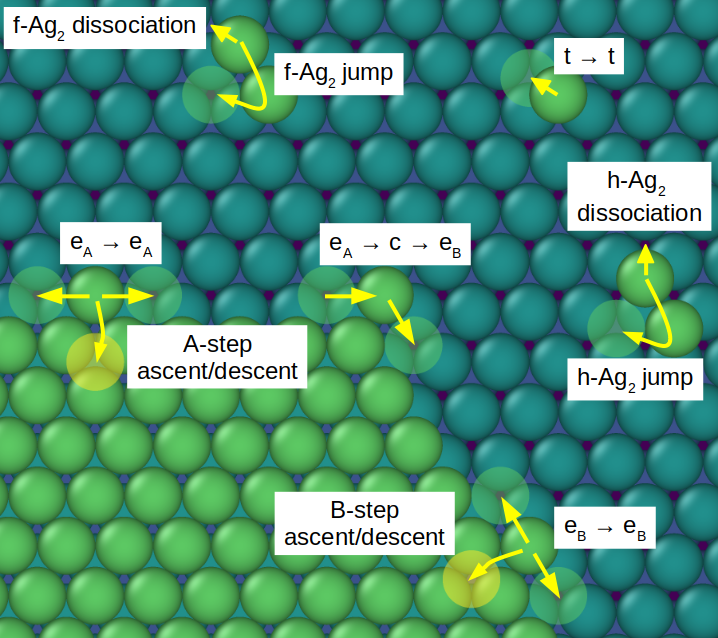}
      \caption{Key processes chosen to be evaluated after the initial training pipeline. Notation: t is terrace, e is edge, c is corner, with f and h marking fcc and hcp sites, respectively. A and B denote the~\hkl(100) and~\hkl(111) microfaceted edges, respectively.}
      \label{fig:processes}
  \end{figure}

  After the initial training, the barrier estimates of the ML model for selected crucial transitions were evaluated. The selected events are illustrated in Fig.~\ref{fig:processes}, and the associated barriers are tabulated in Tab.~\ref{tab:barriers} with comparisons to NEB and literature values. In some cases where multiple different pathways correspond to the same overall event, we tabulate the lowest barrier among them. The pathway with the lowest barrier, i.e., the highest rate, typically dominates (limits) the transitions between these states.
  
  For the case of an Ag dimer, the lowest available barrier corresponds to the ``intracell''~\cite{marinica2004influence}~60\textdegree\ rotation about the central hollow (hcp dimer) or top (fcc dimer) site by exchange, giving no center-of-mass mobility to the dimer as a whole. Thus, to assess the mobility of the dimer in this model, an \emph{effective} dimer diffusion barrier was obtained by simulating the dimer diffusion at temperatures~100--500\,K and fitting
  \begin{equation}
      \label{eq:MSD}
      \langle \left(r_t - r_0\right)^2 \rangle = 4Dt,
  \end{equation}
  where $r_0$ is the initial position of the dimer and $r_t$ is its position at time $t$, and $D$ is the diffusivity
  \begin{equation}
      \label{eq:diffusivity}
      D \propto \exp\left(\frac{-E_m}{k_\mathrm{B}T}\right)
  \end{equation}
  
\begin{figure}
      \centering
      \input{barriers.tex}
      \caption{Correlation between selected NEB calculated and ML estimated barriers from Tab.~\ref{tab:barriers}.}
      \label{fig:barriers}
  \end{figure}
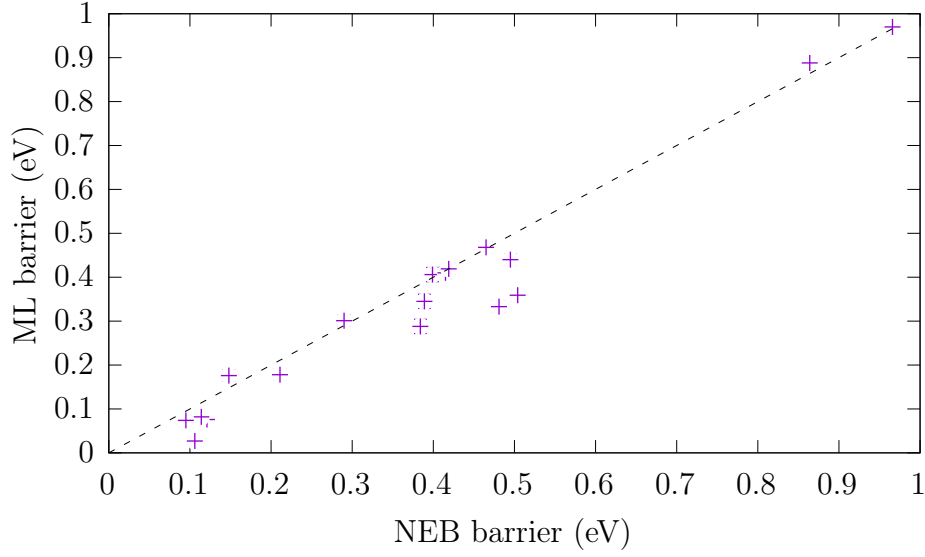

  A correlation plot of the ML barriers vs. NEB is shown in Fig.~\ref{fig:barriers}. Additional pathways with higher barriers exist for most processes, most significant of which is the corner-to-A-edge hop directly to the fcc site; for this jump, the fcc-hcp-fcc pathway has only~11\,meV lower barrier in the ML model.
  
  \begin{table}[h!t]
      \centering
      \caption{Lowest migration energy barriers of selected key processes by NEB, ML regression, and literature. The ML error estimates are the GPR model standard deviations, except for the \emph{effective dimer diffusion barrier} marked with *; this value was extracted by fitting to mean square displacement simulations (see text for more details). The dimer bond energies are taken as the difference between dissociation and association barriers. See the caption of Fig.~\ref{fig:processes} for visualisation and explanation of the event notation. Literature values: a~\cite{lu2018formation}, b~\cite{kim2007transition,kim2007transitionpathway}, c~\cite{bromann1995interlayer,cox2005temperature,morgenstern1998measurement,henzler1994modes,luo1995spa,rosenfeld1995new}, d~\cite{shen2007ripening}.}
      \label{tab:barriers}
      \begin{tabular}{llll}
        \toprule
        \multicolumn{4}{c}{Activation energies (eV)}                                                                                                       \\
        \midrule
        Event                                   & NEB   & ML  & Literature                                                                                 \\
        \midrule
        t$\rightarrow$t (f$\rightarrow$h)       & 0.121 & 0.076 $\pm$ 0.011 & 0.066$^\mathrm{a}$, 0.063$^\mathrm{b}$, 0.05--0.18$^\mathrm{c}$              \\
        t$\rightarrow$t (h$\rightarrow$f)       & 0.106 & 0.027 $\pm$ 0.012 &                                                                              \\
        f-Ag$_2$ jump                           & 0.114 & 0.082 $\pm$ 0.011 & 0.108$^\mathrm{b}$                                                           \\
        h-Ag$_2$ jump                           & 0.095 & 0.074 $\pm$ 0.011 &                                                                              \\
        Ag$_2$ effective*                       &       & 0.179 $\pm$ 0.009 &                                                                              \\
        f-Ag$_2$ dissoc.                        & 0.504 & 0.359 $\pm$ 0.012 &                                                                              \\
        h-Ag$_2$ dissoc.                        & 0.481 & 0.333 $\pm$ 0.012 &                                                                              \\
        f-Ag$_2$ bond                           & 0.389 & 0.345 $\pm$ 0.016 & 0.19--0.24$^\mathrm{d}$                                                      \\
        h-Ag$_2$ bond                           & 0.384 & 0.288 $\pm$ 0.016 &                                                                              \\
        e$_\mathrm{A}\rightarrow$e$_\mathrm{A}$ & 0.399 & 0.406 $\pm$ 0.016 & 0.259$^\mathrm{a}$, 0.258$^\mathrm{b}$, 0.275$^\mathrm{b}$                   \\
        e$_\mathrm{B}\rightarrow$e$_\mathrm{B}$ & 0.465 & 0.468 $\pm$ 0.012 & 0.306$^\mathrm{a}$, 0.302$^\mathrm{b}$, 0.310$^\mathrm{b}$                   \\
        A-step descent                          & 0.415 & 0.410 $\pm$ 0.013 & 0.436$^\mathrm{a}$, 0.333$^\mathrm{b}$, 0.22$^\mathrm{c}$, 0.23$^\mathrm{c}$ \\
        B-step descent                          & 0.290 & 0.301 $\pm$ 0.013 & 0.428$^\mathrm{a}$, 0.332$^\mathrm{b}$                                       \\
        A-step ascent                           & 0.966 & 0.970 $\pm$ 0.011 & 1.066$^\mathrm{a}$, 0.821$^\mathrm{b}$                                       \\
        B-step ascent                           & 0.864 & 0.888 $\pm$ 0.010 & 1.073$^\mathrm{a}$, 0.825$^\mathrm{b}$                                       \\
        e$_\mathrm{A}\rightarrow$c              & 0.419 & 0.419 $\pm$ 0.012 & 0.327$^\mathrm{a}$                                                           \\
        e$_\mathrm{B}\rightarrow$c              & 0.495 & 0.440 $\pm$ 0.013 & 0.385$^\mathrm{a}$, 0.360$^\mathrm{b}$                                       \\
        c$\rightarrow$e$_\mathrm{A}$            & 0.148 & 0.176 $\pm$ 0.013 & 0.065$^\mathrm{a}$, 0.072$^\mathrm{b}$                                       \\
        c$\rightarrow$e$_\mathrm{B}$            & 0.211 & 0.178 $\pm$ 0.012 & 0.109$^\mathrm{a}$, 0.098$^\mathrm{b}$                                       \\
        \bottomrule
      \end{tabular}
  \end{table}
  The model predicts overall the migration barriers for the key processes quite accurately. Terrace adatom and dimer migration barriers as well as the dimer bond energies are slightly underestimated by ML. This underestimation may not be of crucial importance, as even lower values exist in the literature for most cases. Edge diffusion and step crossing are captured well. Dimer dissociation barriers, as well as dimer bond energies, are underestimated by the ML model, but the relative stability of fcc and hcp dimers is preserved.
  
\begin{figure}
      \centering
      \begin{subfigure}{0.32\linewidth}
        \includegraphics[width=\linewidth]{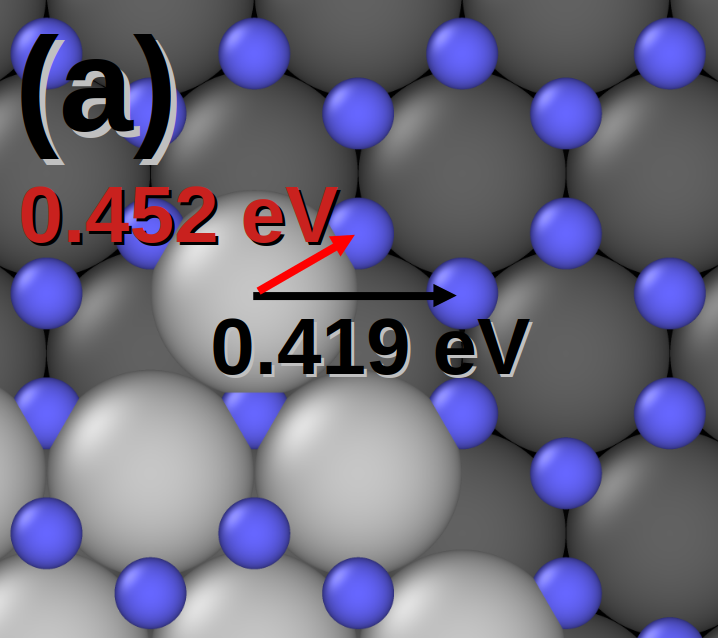}
      \end{subfigure}
      \begin{subfigure}{0.32\linewidth}
        \includegraphics[width=\linewidth]{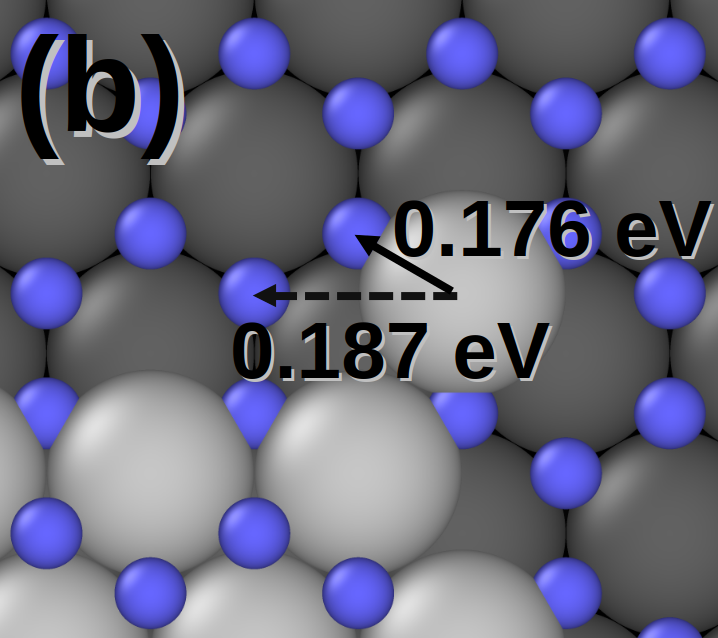}
      \end{subfigure}
      \begin{subfigure}{0.32\linewidth}
        \includegraphics[width=\linewidth]{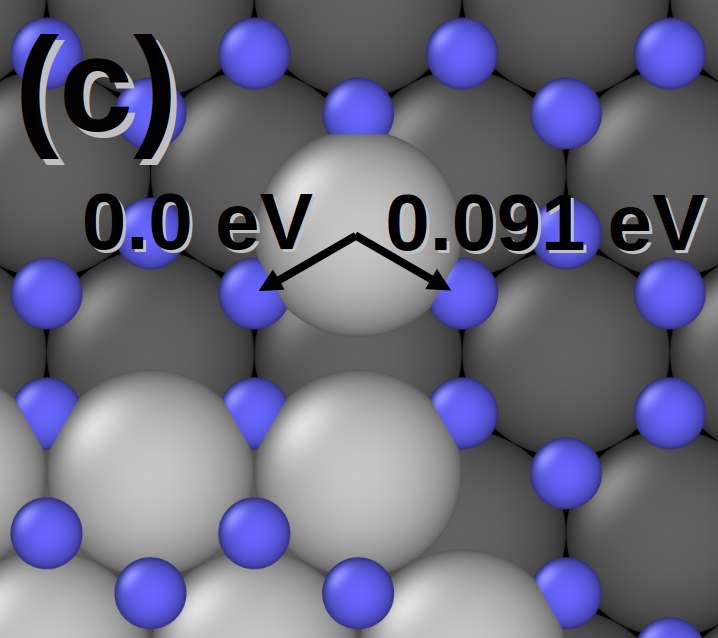}
      \end{subfigure}
      \caption{Corner-A-edge motion with barriers predicted by the ML model. \emph{(a)} Adatom at A-edge may jump directly to the corner site by barrier of~0.419\,eV; jump to the adjacent hcp site is forbidden because the reverse jump is spontaneous (the hcp site is metastable with a local potential energy plateau) \emph{(b)} Adatom at the corner site is allowed to jump to the hcp site because the reverse barrier is finite. Direct jump to the A-edge is normally occluded by the corner-hcp jump, but if the adatom's previous location was at the A-edge, the reverse jump is allowed for algorithmic efficiency reasons. \emph{(c)} Adatom at the hcp site may jump to the A-edge by zero barrier or to the corner by a barrier of~0.091\,eV.}
      \label{fig:corner_detail}
  \end{figure}

  Likewise, the edge-to-corner barriers are reproduced in the correct order, although with somewhat reduced asymmetry. However, for the corner-to-edge hop the lowest-barrier asymmetry is virtually lost. This weakens the corner-crossing anisotropy, which is the primary driving mechanism for preferential bounding of triangular islands~\cite{michely2004islands}, but it is partially recovered by the second pathway from the corner directly to the 1nn fcc site at A-edge. The intermediate hcp site is metastable with zero barrier (spontaneous) jump to the fcc site of the A-edge; thus, the jump from the A-edge towards the corner must take place to the nearest stable site at the corner, in which case the reverse process is exceptionally marked as allowed even though it would normally be occluded by the shorter fcc-hcp hop (see Fig~\ref{fig:corner_detail} for illustration).
  
  Note that the reported ML uncertainties are only slightly higher than the regularisation constant~0.01\,eV for each case. The reason is that each process is represented in the training data---regularisation is implemented in the GAP library so that the uncertainty estimate is the root sum squared of the regularisation constant and the native GPR standard deviation. The native standard deviation is minimal near the training data points.

\subsection{Simulations of submonolayer deposition}

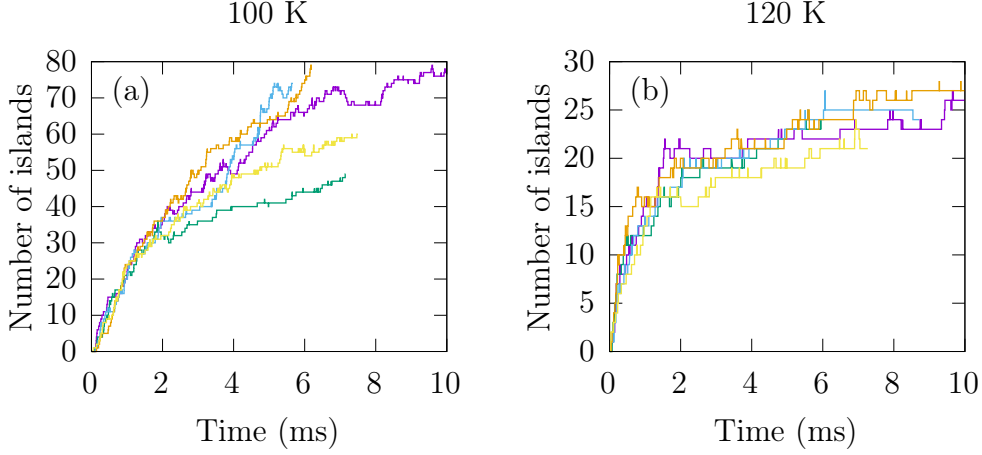
\begin{figure}
      \centering
      \begin{subfigure}{0.49\linewidth}
        \input{islands_vs_time_100K.tex}
      \end{subfigure}
      \begin{subfigure}{0.49\linewidth}
        \input{islands_vs_time_120K.tex}
      \end{subfigure}
      \caption{Evolution of adatom island number vs. simulated time in five simulations at (a)~100\,K and (b)~120\,K.}
      \label{fig:islands_vs_time}
  \end{figure}

  Using the pre-trained ML model as a starting point, statistics were collected by performing five deposition simulations, each using a different random seed at a given temperature. The simulations were conducted at~100\,K, 120\,K, and at temperatures from~150\,K to~500\,K in increments of~50\,K. The same attempt frequency $\nu=1.25\cdot 10^{12}$\,Hz was used for all processes. To reach the saturation island density~\cite{michely2004islands}, we aimed to run all simulations until the coverage of~0.1 monolayers, corresponding to simulated time of~10\,ms at the deposition rate of~10 monolayers/s, but at~100\,K and~120\,K the processes developed very slowly and we were in most cases able to reach only~5\,ms which required approximately~10 weeks of computation time; however, this was sufficient for the island density to saturate, as shown in Fig.~\ref{fig:islands_vs_time}. At temperatures above~150\,K, nearly all simulations reached~10\,ms within two weeks of computation wall time. The final ML models fitted on the merged training sets from all simulations in this article are available at the Fairdata Etsin repository~\cite{kimari2026datasets}, to be utilized and refined in subsequent works. The repository also contains the training data in a compact form to facilitate usage as a test case for machine learning method development.
  
  The reason that the low-temperature simulations proceeded slower is the following. The simulations follow the self-learning algorithm described in Sec.~\ref{sec:methods}, i.e., the ML models in each simulation collects new barrier data and retrain themselves independently. Whenever the ML model reports an uncertainty that exceeds the tolerance of~0.05\,eV, a NEB calculation is launched and the newly obtained barrier is added to the training data. As more data are gathered, the uncertainty tolerance is exceeded less frequently, and the total number of NEB calculations tends to saturate. At~350\,K and~400\,K, the total of new NEB calculations was the lowest, only a few hundred. At even higher temperatures, the processes with higher barriers become possible, momentarily leading to more exotic LAEs and requiring some thousands of new NEB calculations in total. At lower temperatures, on the other hand, the behaviour of the system approaches diffusion-limited aggregation that favors formation of rough island edges, and again thousands or even tens of thousands of new NEB calculations were needed by the end of the simulations. In the low temperature scenario ($T$<150\,K), initially the NEB calculations are the efficiency bottleneck, but as they become less frequent, the construction of the LAEs, the formation of the SOAP descriptors and fetching predictions from the GPR model surpass the NEB calculations and the basic KMC steps in the usage of computation time.
  
\begin{figure}
    \centering
    \begin{subfigure}{0.32\linewidth}
      \centering
      \includegraphics[width=\linewidth]{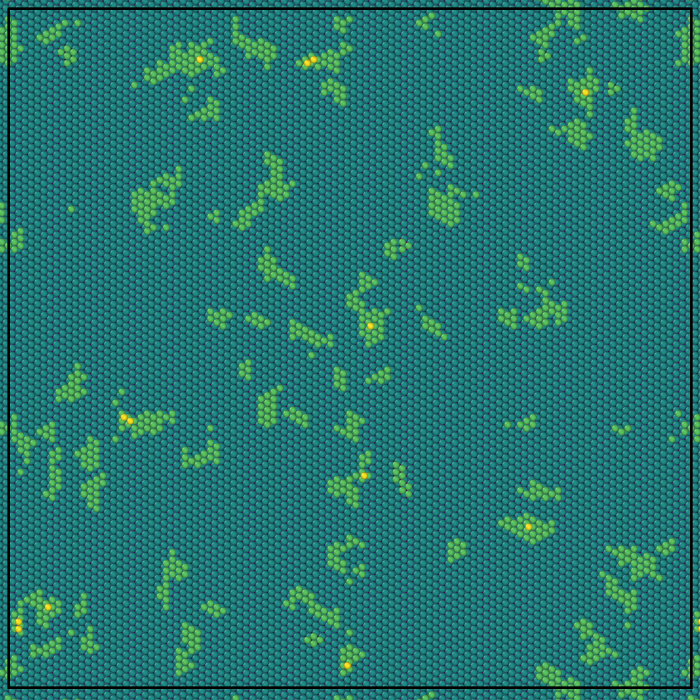}
      \caption{100\,K}
    \end{subfigure}
    \begin{subfigure}{0.32\linewidth}
      \centering
      \includegraphics[width=\linewidth]{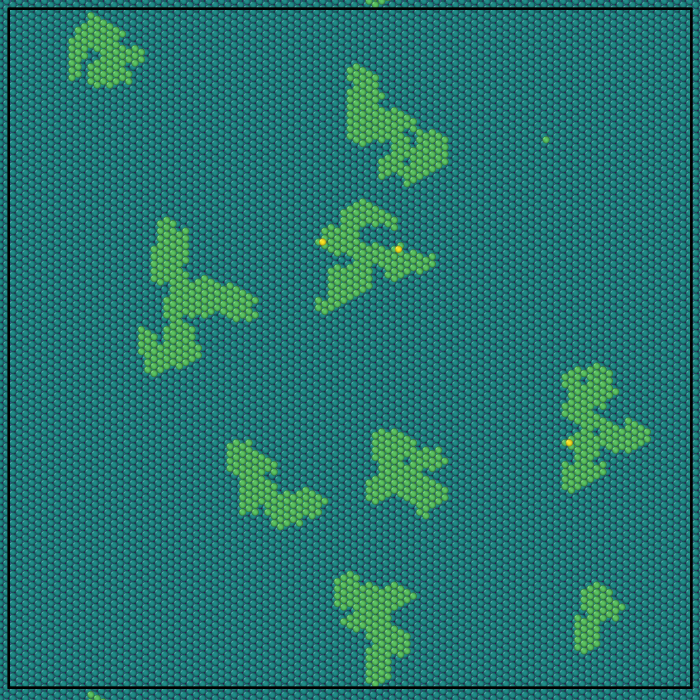}
      \caption{150\,K}
    \end{subfigure}
    \begin{subfigure}{0.32\linewidth}
      \centering
      \includegraphics[width=\linewidth]{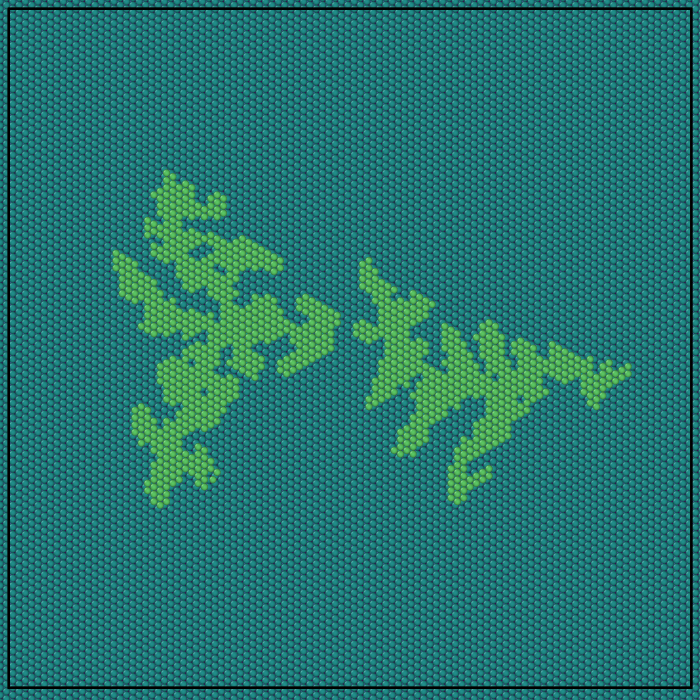}
      \caption{200\,K}
    \end{subfigure}
    \begin{subfigure}{0.32\linewidth}
      \centering
      \includegraphics[width=\linewidth]{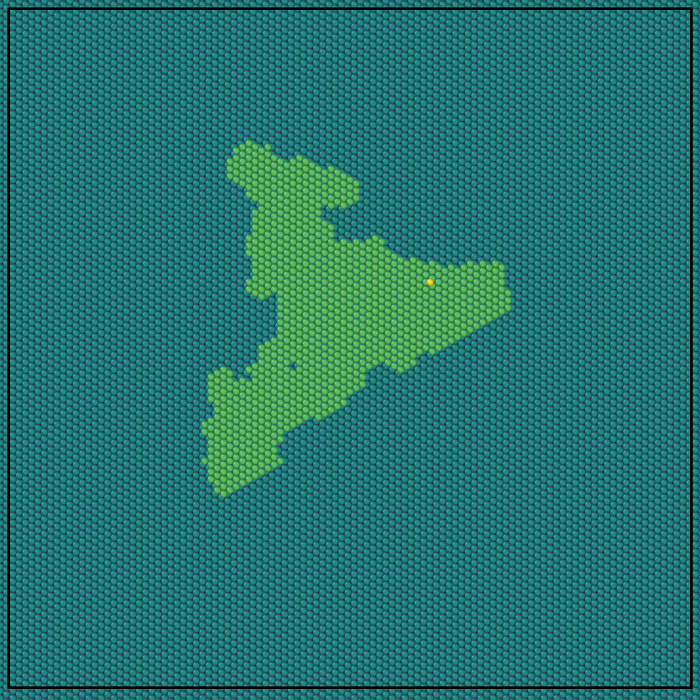}
      \caption{300\,K}
    \end{subfigure}
    \begin{subfigure}{0.32\linewidth}
      \centering
      \includegraphics[width=\linewidth]{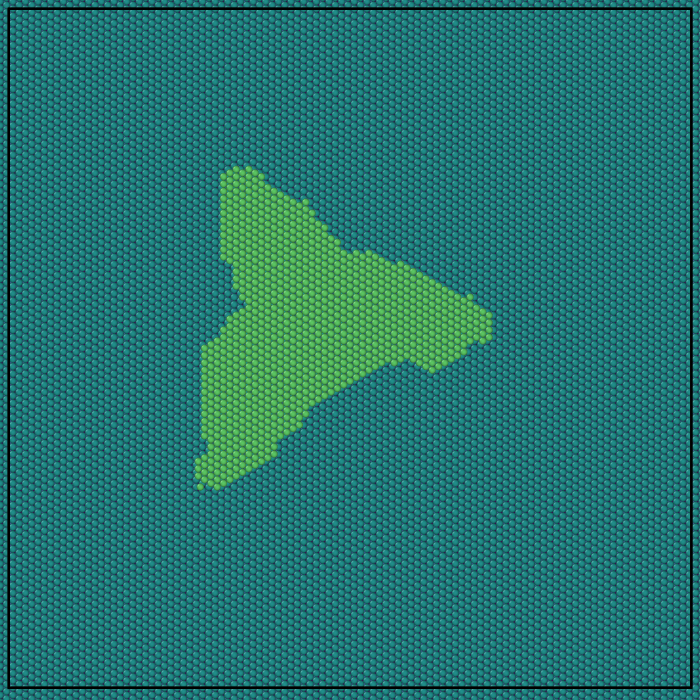}
      \caption{350\,K}
    \end{subfigure}
    \begin{subfigure}{0.32\linewidth}
      \centering
      \includegraphics[width=\linewidth]{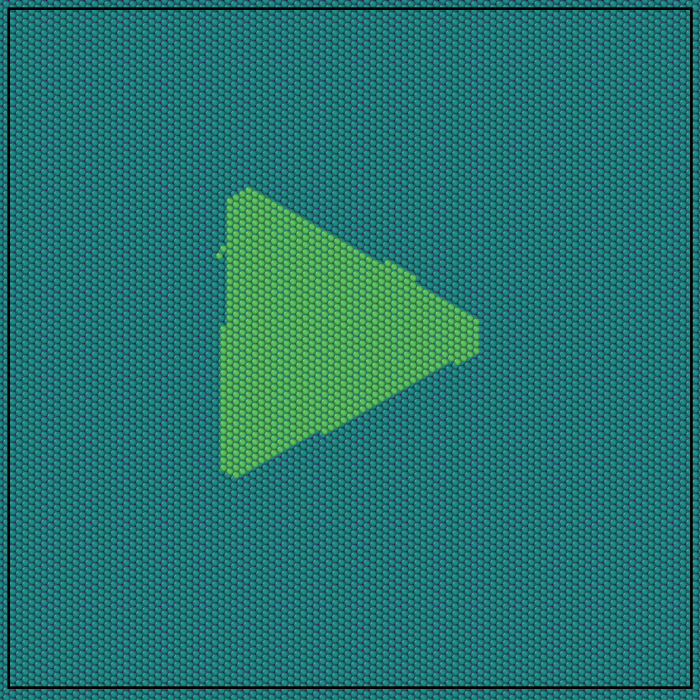}
      \caption{500\,K}
    \end{subfigure}
    \caption{Representative examples of island shapes as a function of temperature at the end of the simulation runs after deposition of~0.1 monolayers.}
    \label{fig:shapes}
  \end{figure}

  Representative examples of the shapes of the islands at the end of the simulations at different temperatures are shown in Figs.~\ref{fig:shapes}. The shape evolution is qualitatively consistent with available experimental and computational data ranging from numerous small clusters at temperatures <200\,K to large dendritic islands at~200--250\,K, and finally to compact and triangular islands at higher temperatures.~\cite{cox2005temperature} The model reproduces the concave edge features at intermediate temperatures (350\,K in our case) between fractal-dendritic and triangles with straight edges. The triangular islands are bounded by B-edges, driven by corner-crossing anisotropy---in the near-absence of lowest-barrier asymmetry for corner-to-edge jumps (see Tab~\ref{tab:barriers}), we attribute this behaviour to the additional pathway from the corner directly to the fcc site towards the A-edge. At~0.1 monolayer coverage, apart from a few dimers at very low temperatures, we do not observe significant second-layer nucleation.

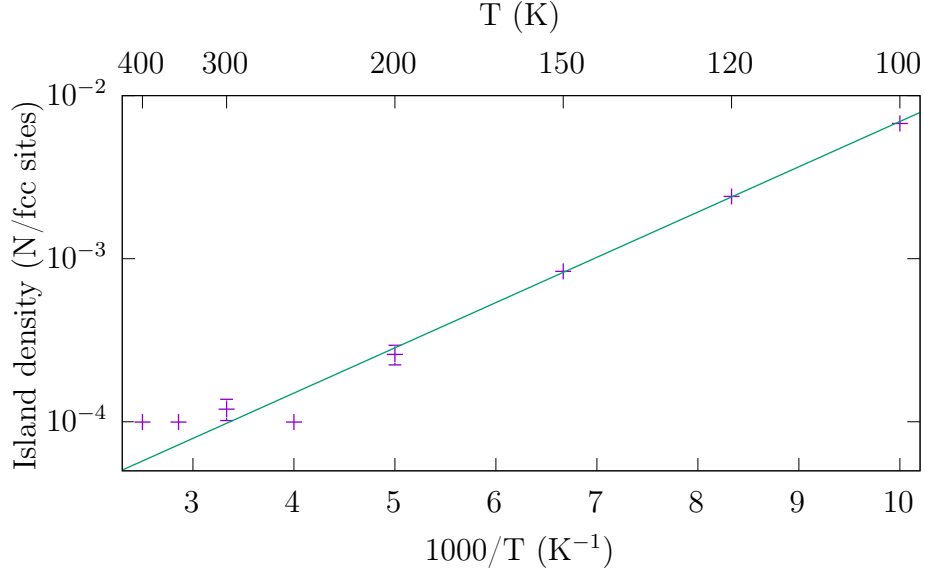
\begin{figure}
      \centering
      \input{island_density.tex}
      \caption{Island density as a function of inverse temperature. The solid line corresponds to a fit of Eq.~\eqref{eq:island_barriers} in the 100--200\,K range. At temperatures above~200\,K, the simulation cell size limits the island density (only one island fits in the cell).}
      \label{fig:island_density}
  \end{figure}

  Along with with the temperature-dependent island shape evolution, we investigated the scaling of the island density as a function of temperature. As temperature increases from~100\,K, the island density decreases from~0.007/fcc site until it saturates above~200\,K to~10$^{-4}$/fcc site, which corresponds to one island per simulation cell (see Fig.~\ref{fig:island_density}). Island density follows an exponential behaviour determined by the activation energy barrier of \emph{mobile units}. If only monomers are mobile, with dimers and larger islands immobile and stable, the island density is given by~\cite{michely2004islands}:
  \begin{equation}
      \label{eq:island_diffusivity}
      \hat N \propto \left(\frac{1}{D_1}\right)^{1/3},
  \end{equation}
  where $D_1$ is the monomer diffusivity. In the presence of mobile but stable dimers, the island density follows~\cite{michely2004islands}:
  \begin{equation}
      \label{eq:island_diffusivity_with_dimers}
      \hat N \propto \left(\frac{1}{D_1D_2}\right)^{1/5},
  \end{equation}
  where $D_2$ is the effective diffusivity of the dimer. According to the dimer migration and dissociation barriers in our model (see Tab.~\ref{tab:barriers}), dimer mobility begins roughly at~110\,K, while dimer dissociation will not onset until~210\,K; island density is expected to follow Eq.~\eqref{eq:island_diffusivity_with_dimers}. To obtain a reverse estimate for the sum of the mobile unit migration barriers, we substitute Eq.~\eqref{eq:diffusivity}:
  \begin{equation}
      \label{eq:island_barriers}
      \hat N \propto \left(\frac{E_\mathrm{m1}+E_\mathrm{m2}}{5k_\mathrm{B}T}\right)
  \end{equation}
  We fit Eq.~\eqref{eq:island_barriers} to the island density data with the result~0.275~$\pm$~0.006\,eV for the sum of barriers. The fitted effective migration barrier is slightly higher but agrees within error margins with the value given by the ML model for the sum of the limiting barrier for the monomer and the effective barrier for the dimer, equal to~0.255~$\pm$~0.015\,eV. This internal consistency confirms that the implementation of diffusion kinetics in our model is integrous. (For comparison, fitting Eq.~\eqref{eq:island_diffusivity} to the simulated island densities yields an estimate of~0.1653\,eV for the monomer migration barrier, much higher than the rate-limiting barrier of~0.074\,eV used in the simulations.)

  The overall agreement of our data with respect to the temperature-dependent island density and shape evolution with available experimental and computational data as well as the theoretical framework pertaining to far-from-equilibrium epitaxial growth shows that our model can correctly capture the dynamics and atomic-scale processes that govern deposition of Ag on Ag~\hkl{111} in the sub-monolayer regime.
  
\section{Summary and outlook}
\label{sec:conclusions}

  We have developed a machine-learning~(ML) augmented kinetic Monte Carlo~(KMC) model for metal surface deposition and diffusion, that self-parameterizes given an interatomic potential energy function. The model operates on a dynamic refined lattice: starting from a regular fcc~\hkl{111} surface, both fcc and hcp sites are available for adatom deposition and jumps, with new sites generated as necessary for new layers and arbitrarily oriented facets. We have demonstrated the performance of the model in the case of submonolayer deposition of Ag on Ag~\hkl{111}. The self-parameterization scheme learns the migration barrier distribution of the underlying potential energy function reasonably well, and the simulations produce island shapes and densities as expected.

  The computational bottlenecks of our approach are, at first, the nudged elastic band calculations that create the training set for the ML model and, subsequently, the generation of the ML descriptors and fetching the parameter predictions. The efficiency may be improved by further optimizations of the descriptor or by exploring other options for the ML model besides the Gaussian process regression that was selected for this proof-of-concept work---any model that can estimate its own uncertainty is a possible substitute into the workflow.

  The simulation tool presented herein readily supports multielemental systems, given a suitable potential energy function. Implementing other substrate orientations or even crystal types in addition to the currently available fcc~\hkl{111} requires no fundamental changes in the codebase. The model may also be applied to (vacancy assisted) diffusion in bulk materials.

\section*{CRediT authorship contribution statement}

  \textbf{Jyri Kimari:} Writing -- Original Draft, Conceptualization, Methodology, Software, Investigation, Formal analysis, Data Curation, Visualization. \textbf{Flyura Djurabekova:} Writing -- Review \& Editing, Conceptualization. \textbf{Kostas Sarakinos:} Writing -- Review \& Editing, Conceptualization, Resources, Project administration, Funding acquisition.

\section*{Declaration of competing interest}

  The authors declare that they have no known competing financial interests or personal relationships that could have appeared to influence the work reported in this paper.
  
\section*{Acknowledgments}
\label{sec:acknowledgments}

  We highly acknowledge Professor Teemu Roos and Dr Jussi Määttä for valuable discussions related to the development of the ML-KMC algorithm. The computations were enabled by resources provided by the National Academic Infrastructure for Supercomputing in Sweden (NAISS), partially funded by the Swedish Research Council through grant agreement no.~2022-06725. The authors acknowledge CSC---IT Center for Science, Finland, for computational resources. The authors wish to thank the Finnish Computing Competence Infrastructure (FCCI) for supporting this project with computational and data storage resources. JK and KS acknowledge financial support by the Swedish Research Council (Grant No. VR-2021-04113), the ÅForsk foundation (Grant No. 22-150), and the Research Council of Finland (Grant No. 372528). FD acknowledges the SPATEC project (Grant No. 349690).

\bibliography{jyri_kimari.bib}

\end{document}

%% file: barriers.tex
\begingroup
  \makeatletter
  \providecommand\color[2][]{%
    \GenericError{(gnuplot) \space\space\space\@spaces}{%
      Package color not loaded in conjunction with
      terminal option `colourtext'%
    }{See the gnuplot documentation for explanation.%
    }{Either use 'blacktext' in gnuplot or load the package
      color.sty in LaTeX.}%
    \renewcommand\color[2][]{}%
  }%
  \providecommand\includegraphics[2][]{%
    \GenericError{(gnuplot) \space\space\space\@spaces}{%
      Package graphicx or graphics not loaded%
    }{See the gnuplot documentation for explanation.%
    }{The gnuplot epslatex terminal needs graphicx.sty or graphics.sty.}%
    \renewcommand\includegraphics[2][]{}%
  }%
  \providecommand\rotatebox[2]{#2}%
  \@ifundefined{ifGPcolor}{%
    \newif\ifGPcolor
    \GPcolortrue
  }{}%
  \@ifundefined{ifGPblacktext}{%
    \newif\ifGPblacktext
    \GPblacktexttrue
  }{}%
  \let\gplgaddtomacro\g@addto@macro
  \gdef\gplbacktext{}%
  \gdef\gplfronttext{}%
  \makeatother
  \ifGPblacktext
    \def\colorrgb#1{}%
    \def\colorgray#1{}%
  \else
    \ifGPcolor
      \def\colorrgb#1{\color[rgb]{#1}}%
      \def\colorgray#1{\color[gray]{#1}}%
      \expandafter\def\csname LTw\endcsname{\color{white}}%
      \expandafter\def\csname LTb\endcsname{\color{black}}%
      \expandafter\def\csname LTa\endcsname{\color{black}}%
      \expandafter\def\csname LT0\endcsname{\color[rgb]{1,0,0}}%
      \expandafter\def\csname LT1\endcsname{\color[rgb]{0,1,0}}%
      \expandafter\def\csname LT2\endcsname{\color[rgb]{0,0,1}}%
      \expandafter\def\csname LT3\endcsname{\color[rgb]{1,0,1}}%
      \expandafter\def\csname LT4\endcsname{\color[rgb]{0,1,1}}%
      \expandafter\def\csname LT5\endcsname{\color[rgb]{1,1,0}}%
      \expandafter\def\csname LT6\endcsname{\color[rgb]{0,0,0}}%
      \expandafter\def\csname LT7\endcsname{\color[rgb]{1,0.3,0}}%
      \expandafter\def\csname LT8\endcsname{\color[rgb]{0.5,0.5,0.5}}%
    \else
      \def\colorrgb#1{\color{black}}%
      \def\colorgray#1{\color[gray]{#1}}%
      \expandafter\def\csname LTw\endcsname{\color{white}}%
      \expandafter\def\csname LTb\endcsname{\color{black}}%
      \expandafter\def\csname LTa\endcsname{\color{black}}%
      \expandafter\def\csname LT0\endcsname{\color{black}}%
      \expandafter\def\csname LT1\endcsname{\color{black}}%
      \expandafter\def\csname LT2\endcsname{\color{black}}%
      \expandafter\def\csname LT3\endcsname{\color{black}}%
      \expandafter\def\csname LT4\endcsname{\color{black}}%
      \expandafter\def\csname LT5\endcsname{\color{black}}%
      \expandafter\def\csname LT6\endcsname{\color{black}}%
      \expandafter\def\csname LT7\endcsname{\color{black}}%
      \expandafter\def\csname LT8\endcsname{\color{black}}%
    \fi
  \fi
    \setlength{\unitlength}{0.0500bp}%
    \ifx\gptboxheight\undefined%
      \newlength{\gptboxheight}%
      \newlength{\gptboxwidth}%
      \newsavebox{\gptboxtext}%
    \fi%
    \setlength{\fboxrule}{0.5pt}%
    \setlength{\fboxsep}{1pt}%
    \definecolor{tbcol}{rgb}{1,1,1}%
\begin{picture}(7200.00,4320.00)%
    \gplgaddtomacro\gplbacktext{%
      \csname LTb\endcsname
      \put(692,767){\makebox(0,0)[r]{\strut{}$0$}}%
      \csname LTb\endcsname
      \put(692,1096){\makebox(0,0)[r]{\strut{}$0.1$}}%
      \csname LTb\endcsname
      \put(692,1425){\makebox(0,0)[r]{\strut{}$0.2$}}%
      \csname LTb\endcsname
      \put(692,1755){\makebox(0,0)[r]{\strut{}$0.3$}}%
      \csname LTb\endcsname
      \put(692,2084){\makebox(0,0)[r]{\strut{}$0.4$}}%
      \csname LTb\endcsname
      \put(692,2413){\makebox(0,0)[r]{\strut{}$0.5$}}%
      \csname LTb\endcsname
      \put(692,2742){\makebox(0,0)[r]{\strut{}$0.6$}}%
      \csname LTb\endcsname
      \put(692,3072){\makebox(0,0)[r]{\strut{}$0.7$}}%
      \csname LTb\endcsname
      \put(692,3401){\makebox(0,0)[r]{\strut{}$0.8$}}%
      \csname LTb\endcsname
      \put(692,3730){\makebox(0,0)[r]{\strut{}$0.9$}}%
      \csname LTb\endcsname
      \put(692,4060){\makebox(0,0)[r]{\strut{}$1$}}%
      \csname LTb\endcsname
      \put(793,527){\makebox(0,0){\strut{}$0$}}%
      \csname LTb\endcsname
      \put(1401,527){\makebox(0,0){\strut{}$0.1$}}%
      \csname LTb\endcsname
      \put(2010,527){\makebox(0,0){\strut{}$0.2$}}%
      \csname LTb\endcsname
      \put(2618,527){\makebox(0,0){\strut{}$0.3$}}%
      \csname LTb\endcsname
      \put(3227,527){\makebox(0,0){\strut{}$0.4$}}%
      \csname LTb\endcsname
      \put(3835,527){\makebox(0,0){\strut{}$0.5$}}%
      \csname LTb\endcsname
      \put(4444,527){\makebox(0,0){\strut{}$0.6$}}%
      \csname LTb\endcsname
      \put(5052,527){\makebox(0,0){\strut{}$0.7$}}%
      \csname LTb\endcsname
      \put(5660,527){\makebox(0,0){\strut{}$0.8$}}%
      \csname LTb\endcsname
      \put(6269,527){\makebox(0,0){\strut{}$0.9$}}%
      \csname LTb\endcsname
      \put(6877,527){\makebox(0,0){\strut{}$1$}}%
    }%
    \gplgaddtomacro\gplfronttext{%
      \csname LTb\endcsname
      \put(195,2413){\rotatebox{-270.00}{\makebox(0,0){\strut{}ML barrier (eV)}}}%
      \csname LTb\endcsname
      \put(3835,167){\makebox(0,0){\strut{}NEB barrier (eV)}}%
    }%
    \gplbacktext
    \put(0,0){\includegraphics[width={360.00bp},height={216.00bp}]{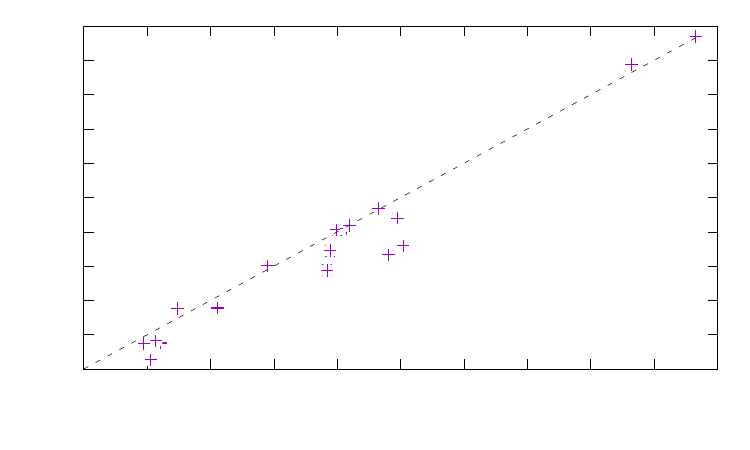}}%
    \gplfronttext
  \end{picture}%
\endgroup

%% file: islands_vs_time_100K.tex
\begingroup
  \makeatletter
  \providecommand\color[2][]{%
    \GenericError{(gnuplot) \space\space\space\@spaces}{%
      Package color not loaded in conjunction with
      terminal option `colourtext'%
    }{See the gnuplot documentation for explanation.%
    }{Either use 'blacktext' in gnuplot or load the package
      color.sty in LaTeX.}%
    \renewcommand\color[2][]{}%
  }%
  \providecommand\includegraphics[2][]{%
    \GenericError{(gnuplot) \space\space\space\@spaces}{%
      Package graphicx or graphics not loaded%
    }{See the gnuplot documentation for explanation.%
    }{The gnuplot epslatex terminal needs graphicx.sty or graphics.sty.}%
    \renewcommand\includegraphics[2][]{}%
  }%
  \providecommand\rotatebox[2]{#2}%
  \@ifundefined{ifGPcolor}{%
    \newif\ifGPcolor
    \GPcolortrue
  }{}%
  \@ifundefined{ifGPblacktext}{%
    \newif\ifGPblacktext
    \GPblacktexttrue
  }{}%
  \let\gplgaddtomacro\g@addto@macro
  \gdef\gplbacktext{}%
  \gdef\gplfronttext{}%
  \makeatother
  \ifGPblacktext
    \def\colorrgb#1{}%
    \def\colorgray#1{}%
  \else
    \ifGPcolor
      \def\colorrgb#1{\color[rgb]{#1}}%
      \def\colorgray#1{\color[gray]{#1}}%
      \expandafter\def\csname LTw\endcsname{\color{white}}%
      \expandafter\def\csname LTb\endcsname{\color{black}}%
      \expandafter\def\csname LTa\endcsname{\color{black}}%
      \expandafter\def\csname LT0\endcsname{\color[rgb]{1,0,0}}%
      \expandafter\def\csname LT1\endcsname{\color[rgb]{0,1,0}}%
      \expandafter\def\csname LT2\endcsname{\color[rgb]{0,0,1}}%
      \expandafter\def\csname LT3\endcsname{\color[rgb]{1,0,1}}%
      \expandafter\def\csname LT4\endcsname{\color[rgb]{0,1,1}}%
      \expandafter\def\csname LT5\endcsname{\color[rgb]{1,1,0}}%
      \expandafter\def\csname LT6\endcsname{\color[rgb]{0,0,0}}%
      \expandafter\def\csname LT7\endcsname{\color[rgb]{1,0.3,0}}%
      \expandafter\def\csname LT8\endcsname{\color[rgb]{0.5,0.5,0.5}}%
    \else
      \def\colorrgb#1{\color{black}}%
      \def\colorgray#1{\color[gray]{#1}}%
      \expandafter\def\csname LTw\endcsname{\color{white}}%
      \expandafter\def\csname LTb\endcsname{\color{black}}%
      \expandafter\def\csname LTa\endcsname{\color{black}}%
      \expandafter\def\csname LT0\endcsname{\color{black}}%
      \expandafter\def\csname LT1\endcsname{\color{black}}%
      \expandafter\def\csname LT2\endcsname{\color{black}}%
      \expandafter\def\csname LT3\endcsname{\color{black}}%
      \expandafter\def\csname LT4\endcsname{\color{black}}%
      \expandafter\def\csname LT5\endcsname{\color{black}}%
      \expandafter\def\csname LT6\endcsname{\color{black}}%
      \expandafter\def\csname LT7\endcsname{\color{black}}%
      \expandafter\def\csname LT8\endcsname{\color{black}}%
    \fi
  \fi
    \setlength{\unitlength}{0.0500bp}%
    \ifx\gptboxheight\undefined%
      \newlength{\gptboxheight}%
      \newlength{\gptboxwidth}%
      \newsavebox{\gptboxtext}%
    \fi%
    \setlength{\fboxrule}{0.5pt}%
    \setlength{\fboxsep}{1pt}%
    \definecolor{tbcol}{rgb}{1,1,1}%
\begin{picture}(3680.00,3680.00)%
    \gplgaddtomacro\gplbacktext{%
      \csname LTb\endcsname
      \put(592,767){\makebox(0,0)[r]{\strut{}$0$}}%
      \csname LTb\endcsname
      \put(592,1038){\makebox(0,0)[r]{\strut{}$10$}}%
      \csname LTb\endcsname
      \put(592,1310){\makebox(0,0)[r]{\strut{}$20$}}%
      \csname LTb\endcsname
      \put(592,1582){\makebox(0,0)[r]{\strut{}$30$}}%
      \csname LTb\endcsname
      \put(592,1853){\makebox(0,0)[r]{\strut{}$40$}}%
      \csname LTb\endcsname
      \put(592,2125){\makebox(0,0)[r]{\strut{}$50$}}%
      \csname LTb\endcsname
      \put(592,2397){\makebox(0,0)[r]{\strut{}$60$}}%
      \csname LTb\endcsname
      \put(592,2669){\makebox(0,0)[r]{\strut{}$70$}}%
      \csname LTb\endcsname
      \put(592,2940){\makebox(0,0)[r]{\strut{}$80$}}%
      \csname LTb\endcsname
      \put(692,527){\makebox(0,0){\strut{}$0$}}%
      \csname LTb\endcsname
      \put(1225,527){\makebox(0,0){\strut{}$2$}}%
      \csname LTb\endcsname
      \put(1758,527){\makebox(0,0){\strut{}$4$}}%
      \csname LTb\endcsname
      \put(2291,527){\makebox(0,0){\strut{}$6$}}%
      \csname LTb\endcsname
      \put(2824,527){\makebox(0,0){\strut{}$8$}}%
      \csname LTb\endcsname
      \put(3357,527){\makebox(0,0){\strut{}$10$}}%
      \csname LTb\endcsname
      \put(852,2723){\makebox(0,0)[l]{\strut{}(a)}}%
    }%
    \gplgaddtomacro\gplfronttext{%
      \csname LTb\endcsname
      \put(195,1853){\rotatebox{-270.00}{\makebox(0,0){\strut{}Number of islands}}}%
      \csname LTb\endcsname
      \put(2025,167){\makebox(0,0){\strut{}Time (ms)}}%
      \csname LTb\endcsname
      \put(2025,3300){\makebox(0,0){\strut{}100 K}}%
    }%
    \gplbacktext
    \put(0,0){\includegraphics[width={184.00bp},height={184.00bp}]{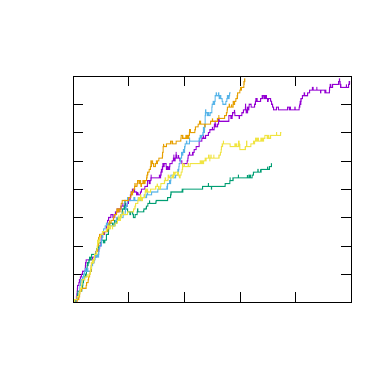}}%
    \gplfronttext
  \end{picture}%
\endgroup

%% file: islands_vs_time_120K.tex
\begingroup
  \makeatletter
  \providecommand\color[2][]{%
    \GenericError{(gnuplot) \space\space\space\@spaces}{%
      Package color not loaded in conjunction with
      terminal option `colourtext'%
    }{See the gnuplot documentation for explanation.%
    }{Either use 'blacktext' in gnuplot or load the package
      color.sty in LaTeX.}%
    \renewcommand\color[2][]{}%
  }%
  \providecommand\includegraphics[2][]{%
    \GenericError{(gnuplot) \space\space\space\@spaces}{%
      Package graphicx or graphics not loaded%
    }{See the gnuplot documentation for explanation.%
    }{The gnuplot epslatex terminal needs graphicx.sty or graphics.sty.}%
    \renewcommand\includegraphics[2][]{}%
  }%
  \providecommand\rotatebox[2]{#2}%
  \@ifundefined{ifGPcolor}{%
    \newif\ifGPcolor
    \GPcolortrue
  }{}%
  \@ifundefined{ifGPblacktext}{%
    \newif\ifGPblacktext
    \GPblacktexttrue
  }{}%
  \let\gplgaddtomacro\g@addto@macro
  \gdef\gplbacktext{}%
  \gdef\gplfronttext{}%
  \makeatother
  \ifGPblacktext
    \def\colorrgb#1{}%
    \def\colorgray#1{}%
  \else
    \ifGPcolor
      \def\colorrgb#1{\color[rgb]{#1}}%
      \def\colorgray#1{\color[gray]{#1}}%
      \expandafter\def\csname LTw\endcsname{\color{white}}%
      \expandafter\def\csname LTb\endcsname{\color{black}}%
      \expandafter\def\csname LTa\endcsname{\color{black}}%
      \expandafter\def\csname LT0\endcsname{\color[rgb]{1,0,0}}%
      \expandafter\def\csname LT1\endcsname{\color[rgb]{0,1,0}}%
      \expandafter\def\csname LT2\endcsname{\color[rgb]{0,0,1}}%
      \expandafter\def\csname LT3\endcsname{\color[rgb]{1,0,1}}%
      \expandafter\def\csname LT4\endcsname{\color[rgb]{0,1,1}}%
      \expandafter\def\csname LT5\endcsname{\color[rgb]{1,1,0}}%
      \expandafter\def\csname LT6\endcsname{\color[rgb]{0,0,0}}%
      \expandafter\def\csname LT7\endcsname{\color[rgb]{1,0.3,0}}%
      \expandafter\def\csname LT8\endcsname{\color[rgb]{0.5,0.5,0.5}}%
    \else
      \def\colorrgb#1{\color{black}}%
      \def\colorgray#1{\color[gray]{#1}}%
      \expandafter\def\csname LTw\endcsname{\color{white}}%
      \expandafter\def\csname LTb\endcsname{\color{black}}%
      \expandafter\def\csname LTa\endcsname{\color{black}}%
      \expandafter\def\csname LT0\endcsname{\color{black}}%
      \expandafter\def\csname LT1\endcsname{\color{black}}%
      \expandafter\def\csname LT2\endcsname{\color{black}}%
      \expandafter\def\csname LT3\endcsname{\color{black}}%
      \expandafter\def\csname LT4\endcsname{\color{black}}%
      \expandafter\def\csname LT5\endcsname{\color{black}}%
      \expandafter\def\csname LT6\endcsname{\color{black}}%
      \expandafter\def\csname LT7\endcsname{\color{black}}%
      \expandafter\def\csname LT8\endcsname{\color{black}}%
    \fi
  \fi
    \setlength{\unitlength}{0.0500bp}%
    \ifx\gptboxheight\undefined%
      \newlength{\gptboxheight}%
      \newlength{\gptboxwidth}%
      \newsavebox{\gptboxtext}%
    \fi%
    \setlength{\fboxrule}{0.5pt}%
    \setlength{\fboxsep}{1pt}%
    \definecolor{tbcol}{rgb}{1,1,1}%
\begin{picture}(3680.00,3680.00)%
    \gplgaddtomacro\gplbacktext{%
      \csname LTb\endcsname
      \put(592,767){\makebox(0,0)[r]{\strut{}$0$}}%
      \csname LTb\endcsname
      \put(592,1129){\makebox(0,0)[r]{\strut{}$5$}}%
      \csname LTb\endcsname
      \put(592,1491){\makebox(0,0)[r]{\strut{}$10$}}%
      \csname LTb\endcsname
      \put(592,1853){\makebox(0,0)[r]{\strut{}$15$}}%
      \csname LTb\endcsname
      \put(592,2216){\makebox(0,0)[r]{\strut{}$20$}}%
      \csname LTb\endcsname
      \put(592,2578){\makebox(0,0)[r]{\strut{}$25$}}%
      \csname LTb\endcsname
      \put(592,2940){\makebox(0,0)[r]{\strut{}$30$}}%
      \csname LTb\endcsname
      \put(692,527){\makebox(0,0){\strut{}$0$}}%
      \csname LTb\endcsname
      \put(1225,527){\makebox(0,0){\strut{}$2$}}%
      \csname LTb\endcsname
      \put(1758,527){\makebox(0,0){\strut{}$4$}}%
      \csname LTb\endcsname
      \put(2291,527){\makebox(0,0){\strut{}$6$}}%
      \csname LTb\endcsname
      \put(2824,527){\makebox(0,0){\strut{}$8$}}%
      \csname LTb\endcsname
      \put(3357,527){\makebox(0,0){\strut{}$10$}}%
      \csname LTb\endcsname
      \put(852,2723){\makebox(0,0)[l]{\strut{}(b)}}%
    }%
    \gplgaddtomacro\gplfronttext{%
      \csname LTb\endcsname
      \put(195,1853){\rotatebox{-270.00}{\makebox(0,0){\strut{}Number of islands}}}%
      \csname LTb\endcsname
      \put(2025,167){\makebox(0,0){\strut{}Time (ms)}}%
      \csname LTb\endcsname
      \put(2025,3300){\makebox(0,0){\strut{}120 K}}%
    }%
    \gplbacktext
    \put(0,0){\includegraphics[width={184.00bp},height={184.00bp}]{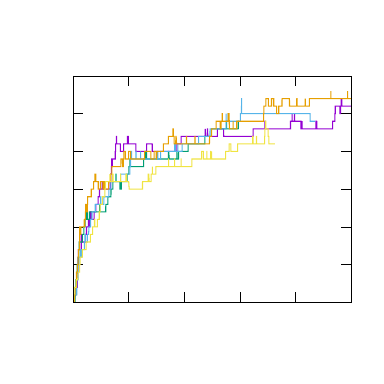}}%
    \gplfronttext
  \end{picture}%
\endgroup

%% file: island_density.tex
\begingroup
  \makeatletter
  \providecommand\color[2][]{%
    \GenericError{(gnuplot) \space\space\space\@spaces}{%
      Package color not loaded in conjunction with
      terminal option `colourtext'%
    }{See the gnuplot documentation for explanation.%
    }{Either use 'blacktext' in gnuplot or load the package
      color.sty in LaTeX.}%
    \renewcommand\color[2][]{}%
  }%
  \providecommand\includegraphics[2][]{%
    \GenericError{(gnuplot) \space\space\space\@spaces}{%
      Package graphicx or graphics not loaded%
    }{See the gnuplot documentation for explanation.%
    }{The gnuplot epslatex terminal needs graphicx.sty or graphics.sty.}%
    \renewcommand\includegraphics[2][]{}%
  }%
  \providecommand\rotatebox[2]{#2}%
  \@ifundefined{ifGPcolor}{%
    \newif\ifGPcolor
    \GPcolortrue
  }{}%
  \@ifundefined{ifGPblacktext}{%
    \newif\ifGPblacktext
    \GPblacktexttrue
  }{}%
  \let\gplgaddtomacro\g@addto@macro
  \gdef\gplbacktext{}%
  \gdef\gplfronttext{}%
  \makeatother
  \ifGPblacktext
    \def\colorrgb#1{}%
    \def\colorgray#1{}%
  \else
    \ifGPcolor
      \def\colorrgb#1{\color[rgb]{#1}}%
      \def\colorgray#1{\color[gray]{#1}}%
      \expandafter\def\csname LTw\endcsname{\color{white}}%
      \expandafter\def\csname LTb\endcsname{\color{black}}%
      \expandafter\def\csname LTa\endcsname{\color{black}}%
      \expandafter\def\csname LT0\endcsname{\color[rgb]{1,0,0}}%
      \expandafter\def\csname LT1\endcsname{\color[rgb]{0,1,0}}%
      \expandafter\def\csname LT2\endcsname{\color[rgb]{0,0,1}}%
      \expandafter\def\csname LT3\endcsname{\color[rgb]{1,0,1}}%
      \expandafter\def\csname LT4\endcsname{\color[rgb]{0,1,1}}%
      \expandafter\def\csname LT5\endcsname{\color[rgb]{1,1,0}}%
      \expandafter\def\csname LT6\endcsname{\color[rgb]{0,0,0}}%
      \expandafter\def\csname LT7\endcsname{\color[rgb]{1,0.3,0}}%
      \expandafter\def\csname LT8\endcsname{\color[rgb]{0.5,0.5,0.5}}%
    \else
      \def\colorrgb#1{\color{black}}%
      \def\colorgray#1{\color[gray]{#1}}%
      \expandafter\def\csname LTw\endcsname{\color{white}}%
      \expandafter\def\csname LTb\endcsname{\color{black}}%
      \expandafter\def\csname LTa\endcsname{\color{black}}%
      \expandafter\def\csname LT0\endcsname{\color{black}}%
      \expandafter\def\csname LT1\endcsname{\color{black}}%
      \expandafter\def\csname LT2\endcsname{\color{black}}%
      \expandafter\def\csname LT3\endcsname{\color{black}}%
      \expandafter\def\csname LT4\endcsname{\color{black}}%
      \expandafter\def\csname LT5\endcsname{\color{black}}%
      \expandafter\def\csname LT6\endcsname{\color{black}}%
      \expandafter\def\csname LT7\endcsname{\color{black}}%
      \expandafter\def\csname LT8\endcsname{\color{black}}%
    \fi
  \fi
    \setlength{\unitlength}{0.0500bp}%
    \ifx\gptboxheight\undefined%
      \newlength{\gptboxheight}%
      \newlength{\gptboxwidth}%
      \newsavebox{\gptboxtext}%
    \fi%
    \setlength{\fboxrule}{0.5pt}%
    \setlength{\fboxsep}{1pt}%
    \definecolor{tbcol}{rgb}{1,1,1}%
\begin{picture}(7200.00,4320.00)%
    \gplgaddtomacro\gplbacktext{%
      \csname LTb\endcsname
      \put(793,1135){\makebox(0,0)[r]{\strut{}$10^{-4}$}}%
      \csname LTb\endcsname
      \put(793,2358){\makebox(0,0)[r]{\strut{}$10^{-3}$}}%
      \csname LTb\endcsname
      \put(793,3580){\makebox(0,0)[r]{\strut{}$10^{-2}$}}%
      \csname LTb\endcsname
      \put(1424,527){\makebox(0,0){\strut{}$3$}}%
      \csname LTb\endcsname
      \put(2181,527){\makebox(0,0){\strut{}$4$}}%
      \csname LTb\endcsname
      \put(2939,527){\makebox(0,0){\strut{}$5$}}%
      \csname LTb\endcsname
      \put(3696,527){\makebox(0,0){\strut{}$6$}}%
      \csname LTb\endcsname
      \put(4454,527){\makebox(0,0){\strut{}$7$}}%
      \csname LTb\endcsname
      \put(5211,527){\makebox(0,0){\strut{}$8$}}%
      \csname LTb\endcsname
      \put(5968,527){\makebox(0,0){\strut{}$9$}}%
      \csname LTb\endcsname
      \put(6726,527){\makebox(0,0){\strut{}$10$}}%
      \csname LTb\endcsname
      \put(1045,3820){\makebox(0,0){\strut{}$400$}}%
      \csname LTb\endcsname
      \put(1676,3820){\makebox(0,0){\strut{}$300$}}%
      \csname LTb\endcsname
      \put(2939,3820){\makebox(0,0){\strut{}$200$}}%
      \csname LTb\endcsname
      \put(4201,3820){\makebox(0,0){\strut{}$150$}}%
      \csname LTb\endcsname
      \put(5464,3820){\makebox(0,0){\strut{}$120$}}%
      \csname LTb\endcsname
      \put(6726,3820){\makebox(0,0){\strut{}$100$}}%
    }%
    \gplgaddtomacro\gplfronttext{%
      \csname LTb\endcsname
      \put(195,2173){\rotatebox{-270.00}{\makebox(0,0){\strut{}Island density (N/fcc sites)}}}%
      \csname LTb\endcsname
      \put(3886,167){\makebox(0,0){\strut{}1000/T (K$^{-1}$)}}%
      \csname LTb\endcsname
      \put(3886,4180){\makebox(0,0){\strut{}T (K)}}%
    }%
    \gplbacktext
    \put(0,0){\includegraphics[width={360.00bp},height={216.00bp}]{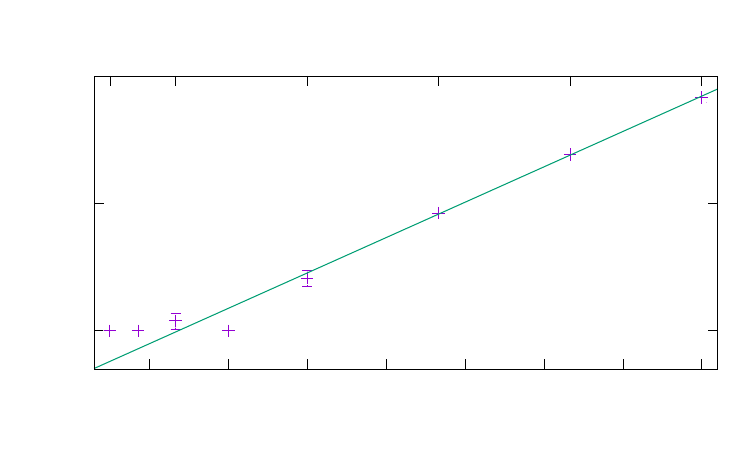}}%
    \gplfronttext
  \end{picture}%
\endgroup